\documentclass[twocolumn,showpacs,preprintnumbers,amsmath,amssymb]{revtex4-1}

\usepackage[dvipdfmx]{graphicx}
\usepackage{dcolumn}
\usepackage{bm}

\begin{document}

\title{
Spectrum of itinerant fractional excitations in quantum spin ice
}

\author{Masafumi Udagawa$^1$ and Roderich Moessner$^2$}%
\affiliation{%
$^1$Department of Physics, Gakushuin University, Mejiro, Toshima-ku, Tokyo 171-8588, Japan\\
$^2$Max-Planck-Institut f\"{u}r Physik komplexer Systeme, 01187 Dresden, Germany
}%

\date{\today}

\begin{abstract}
We study the quantum dynamics of fractional excitations in quantum spin ice. 
We focus on the density of states in the two-monopole
sector, $\rho(\omega)$, as this can be connected to the wavevector-integrated dynamical structure factor accessible 
in neutron scattering experiments. We find that $\rho(\omega)$  exhibits a strikingly characteristic 
singular and asymmetric structure which provides a useful fingerprint for comparison to experiment. 
$\rho(\omega)$ obtained from exact diagonalisation of a finite
cluster agrees  well with that from
the analytical solution of a hopping problem
on a Husimi cactus representing configuration space, but not with the corresponding result on a face-centred cubic lattice, on which
the monopoles move in real space. The main difference between the latter two lies in the inclusion of the emergent gauge field 
degrees of freedom under which the monopoles are charged. This underlines the importance of treating both sets of degrees of freedom 
together, and presents a novel instance of dimensional transmutation.
\end{abstract}

\maketitle
The existence of objects with fractional quantum numbers is by now well established across a range of topologically ordered systems, most notably in quantum spin liquids (QSL)
~\cite{ANDERSON1973153,balents2010balents,Knolle1804.02037}. Their signatures in experiment are not entirely clear, in particular 
to what extent they behave akin to traditional 
 low-energy quasi-particles. There is no simple 
 principle of continuity to a non-interacting limit appeal to, unlike in the case of a Fermi liquid~\cite{anderson2018basic}.
This complicates their theoretical description
 except in the fortunate cases where an exact solution is available, typically at the expense of trading solubility for 
genericity.

The central challenge is to capture the dynamics of the fractional quasiparticle alongside that of the 
emergent gauge field under which it is charged. Mean-field, parton or ad-hoc approaches to achieving this are typically 
not controlled, so that it is e.g.\ not clear what fraction of the excitation spectrum  weakly interacting quasiparticles, even where they exist, 
 occupy in the end. 

Here, we look for qualitative signatures of the quantum dynamics of fractionalised quasiparticles not
in the asymptotic low-energy limit, which may at any rate be hard to probe experimentally,
 but across their full bandwidth. The hope is that gross features and characteristic constraints 
on their exotic properties may thus be rendered accessible.

We focus on quantum spin ice (QSI), one of the simplest and longest-studied QSLs.
Its classical limit, classical 
spin ice (CSI), is well understood~\cite{doi:10.1146/annurev-conmatphys-020911-125058}: 
the macroscopically degenerate ground state of CSI  consists of spin configurations 
 satisfying the ``2-in 2-out" ice rule for all the tetrahedra.
The fractional nature of the elementary excitations already shows up in CSI, 
where a single spin flip out of a ground state decomposes into a pair of tetrahedra (`magnetic monopoles') breaking the ice rule. 
While the monopoles are a priori static in this classical limit,  quantum perturbations turn CSI into QSI, 
enabling these fractional objects to execute coherent quantum motion.

Recently, the coherent motion of quantum monopoles has received increasing attention.
On the experimental side, microwave experiments~\cite{pan2016measure} 
were interpreted in terms of an inertial mass of 
 quantum monopoles  in Yb$_2$Ti$_2$O$_7$, 
  concluding that $m_{\rm eff}\sim2000m_e$, with thermal conductivity measurements~\cite{tokiwa2016tokiwa} 
  suggesting a long mean-free path,
implying highly coherent nature of quantum monopoles.
Inelastic neutrons scattering studies have probed the excitation spectrum of Yb$_2$Ti$_2$O$_7$~\cite{PhysRevX.1.021002,chang2012higgs}, Pr$_2$Zr$_2$O$_7$~\cite{kimura2013quantum}, Pr$_2$Sn$_2$O$_7$~\cite{PhysRevLett.101.227204} and Pr$_2$Hf$_2$O$_7$~\cite{PhysRevB.94.024436,Sibille2018}.
Theoretically, quantum monopoles were explored through a mapping to a Bethe lattice~\cite{PhysRevB.92.100401,petrova2018mesonic}, 
an effective one-spinon theory~\cite{wan2016spinon}, quantum Monte Carlo simulations~\cite{PhysRevLett.120.167202}, and
exact diagonalization of a 2D checkerboard system~\cite{kourtis2016free}.

This work first presents the density of states (DOS) of the two-monopole sector from exact diagonalisation, 
which we argue reliably approximates the  thermodynamic limit.
This DOS turns out to be far from that of a free particle: the coupling to the background gauge field is essential,
leading the DOS to acquire a stronger singularity,  reflected in a 
discontinuous increase at the edge of the wavevector-integrated dynamical structure factor. We capture these features,
which provide characterstic fingerprints for experimental comparisons,
analytically by constructing and solving a hopping problem on a
Husimi cactus. This agrees quantitatively with the numerical results, unlike the qualitatively disagreeing 
analogous treatment of 
monopoles hopping freely on the face-centred cubic lattice of tetrahedra. Further, we show that interactions between monopoles
do not change these results qualitatively, but do have a visible impact in the difference between contractible and non-contractible 
monopole pair configurations. 

{\underline{\it Model:}} We consider a spin-$1/2$ quantum XXZ model on the pyrochlore lattice,
as a minimal model for QSI.
\begin{eqnarray}
\hspace{-0.4cm}
\mathcal{H} \hspace{-0.05cm}= \hspace{-0.05cm}\mathcal{H}_{\rm CSI} + \mathcal{H}_{\rm ex} \hspace{-0.05cm}=\hspace{-0.05cm} \sum_{\langle i,j\rangle}J_zS_i^zS_j^z + \frac{J_{\pm}}{2}(S_i^+S_j^- + S_i^-S_j^+).
\label{eq:Hamiltonian}
\end{eqnarray}
The first term ($\mathcal{H}_{\rm CSI}$) is the antiferromagnetic Ising coupling ($J_z>0$) enforcing the ice rules, and the second term ($\mathcal{H}_{\rm ex}$) induces quantum fluctuation. The spin quantization axes coincide with the local $[111]$ direction.
The Hamiltonian (\ref{eq:Hamiltonian}) serves as a microscopic model for non-Kramers magnets~\cite{PhysRevB.83.094411}, such as the potential QSI compounds Pr$_2$Zr(Sn, Hf)$_2$O$_7$~\cite{kimura2013quantum,PhysRevLett.101.227204,PhysRevB.94.024436,Sibille2018}.

\begin{figure}[h]
\begin{center}
\includegraphics[width=0.5\textwidth]{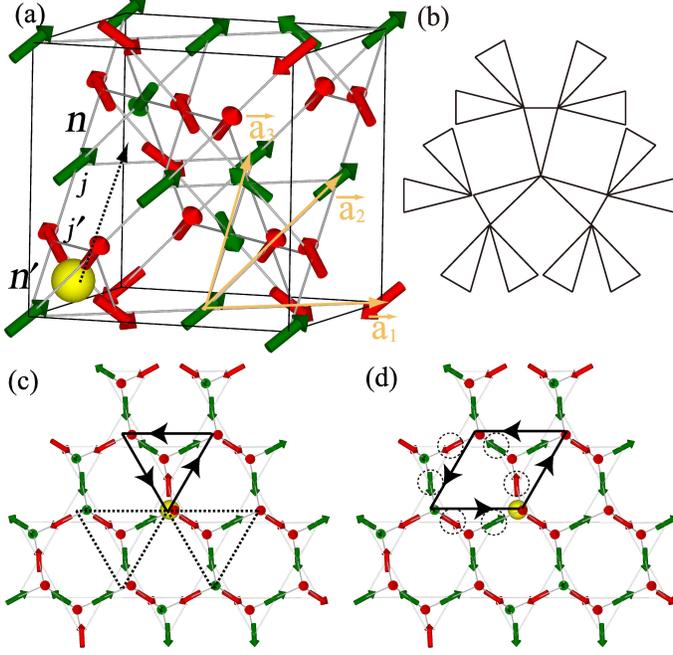}
\end{center}
\caption{\label{Fig1} 
(Color online) (a) Pyrochlore lattice. $\{\vec{a}_1, \vec{a}_2, \vec{a}_3\}$ are the lattice vectors of the FCC lattice of upward tetrahedra. 
A monopole on an upward tetrahedron, $n'$, hops to the neighboring upward tetrahedron, $n$,
by the process $a^{\dag}_n\sigma^x_j\sigma^x_{j'}a_{n'}$ in eq.~(\ref{eq:MonopoleHamiltonian}), by flipping two intervening  
spins $j$ and $j'$.
(b) Schematic picture of the Husimi cactus. 
(c), (d) Pyrochlore lattice seen along the $[111]$ direction. (c) A monopole hops twice following the solid arrows back to the initial tetrahedron. Note that the  spins also return to their initial configuration,  Similar three-step motions are possible for the other two choices of initial hopping directions (dashed lines). These hopping processes imply a mapping to the Husimi cactus, (b). (d) If a monopole comes back to the initial tetrahedron along a larger loop, it goes along with flipping the six spins marked by dashed circles.}
\end{figure}

Here, $S^{z}_{\rm tot}=\sum_iS_i^z$ is a conserved quantity.
We take $J_{\pm}>0$, and consider $J_{\pm}\ll J_z$. For CSI ($J_{\pm}=0$),  the ground states satisfy the ice rule: $\sum_{j\in n}S_j^z=0$ for each tetrahedron, $n$, also implying $S^{z}_{\rm tot}=0$. The first excited level, at energy $J_z$ above the ground state, is also degenerate,  
composed of the states with one pair of monopoles, i.e. two tetrahedra with $\sum_{j\in n}S_j^z=\pm1$. 

The ground-state 
degeneracy is lifted for nonzero $J_{\pm}$, yielding  a ground state splitting of order of $\frac{J_{\pm}^3}{J_z^2}$.
The splitting of the excited level is parametrically larger, of order $J_{\pm}$, suggesting to focus the search 
for signatures of quantum effects on the excitation spectrum rather than the ground state manifold.

The dynamics of a monopole pair can thus be studied by restricting $\mathcal{H}_{\rm ex}$ to the space of two monopoles, enforced
by projection operator $P$,
yielding a simple $\mathcal{H}_{\rm eff}$ in degenerate perturbation theory
\begin{eqnarray}
\hspace{-0.5cm}
\mathcal{H}_{\rm eff} = \frac{J_{\pm}}{2}\sum_{\langle n,n'\rangle}P(a^{\dag}_n\sigma^x_j\sigma^x_{j'}a_{n'} + b^{\dag}_n\sigma^x_j\sigma^x_{j'}b_{n'} + {\rm H.c.})P.
\label{eq:MonopoleHamiltonian}
\end{eqnarray}
We consider the total spin sector $S^{z}_{\rm tot}=1$, and regard the tetrahedron with $\sum_{j\in n}S_j^z=1$ as a monopole.
To describe the two-monopole state, we divide the tetrahedra into two groups, upward and downward [Fig.~\ref{Fig1}(a)] according to their orientations. 
Each group of tetrahedra defines an FCC lattice.
We denote $a^{\dag}_n$ ($b^{\dag}_n$) as creation operator of monopole on an upward (downward) tetrahedron, $n$.
The spin exchange flips a pair of spins, hopping a monopole to a neighboring tetrahedron of the same group
[Fig.~\ref{Fig1}(a)], without disturbing the ice rule for any other tetrahedra.

The dynamical susceptibility, given in terms of the eigenstates $|m\rangle$ of (\ref{eq:Hamiltonian}) with eigenenergies $E_{m}$ as 
\begin{eqnarray}
\hspace{-0.5cm}
\chi_{ij}(\omega) = \sum_{m, m'}\frac{e^{-\beta E_{m'}} - e^{-\beta E_m}}{Z}\frac{\langle m|S^-_i|m'\rangle\langle m'|S^+_j|m\rangle}{\omega - (E_{m'} - E_{m}) + i\delta}.
\label{eq:chiij}
\end{eqnarray}
is connected to the dynamical structure factor, $\mathcal{S}_{\mathbf q}(\omega)$, accessible in inelastic neutron scattering:
the local susceptibility, $\chi_{ii}(\omega)$ is the ${\mathbf q}$-integrated structure factor,
\begin{eqnarray}
\frac{1}{N}\sum_{\mathbf q}\mathcal{S}_{\mathbf q}(\omega) = \frac{\pi}{1-e^{-\beta\omega}}{\rm Im}\ \chi_{ii}(\omega),
\label{eq:observable}
\end{eqnarray}
with $N$ the number of spins.

In the temperature range $\frac{J_{\pm}^3}{J_z^2}\ll T\ll J_z$, the number of excited monopoles is small in equilibrium.  
Quantum coherence is not well developed in the ground state sector, allowing us to replace the summation over $m$ in eq.~(\ref{eq:chiij}) by a simple average over the degenerate CSI ground states for which we set $E_m=0$.
The operation of $S_i^+$, by flipping a spin at site $i$, creates a pair of monopoles, one each on the upward and downward tetrahedra sharing  site $i$.
$|m'\rangle$ and $E_{m'}=J_z + \varepsilon_{m'}$ in eq.~(\ref{eq:chiij}) are obtained from solving $\mathcal{H}_{\rm eff}$, so that 
\begin{align}
\hspace{-0.5cm}
\chi_{ii}(\omega) = -\frac{1}{N_{\rm CSI}}&\sum_{m\in{\rm CSI}}\frac{\langle m|b_{n'}S_i^{-}a_n|m'\rangle\langle m'|a_n^{\dag}S_i^+b_{n'}^{\dag}|m\rangle}{\omega - (J_z + \varepsilon_{m'}) + i\delta},
\label{eq:chiij2}
\end{align}
and the two-monopole density of states,
\begin{align}
\rho(\omega) = \sum_m\delta(\omega - (J_z + \varepsilon_{m})).
\label{eq:DOS}
\end{align}

We thus need all eigenstates of $\mathcal{H}_{\rm eff}$ Hamiltonian (\ref{eq:MonopoleHamiltonian}) in the 
two-monopole Hilbert space, which we construct starting from one spin ice ground state by first flipping an arbitrary spin. 
From this initial state, we generate the other two-monopole states  by considering all possible exchange processes. As far as we have numerically confirmed, such monopole motion is ergodic, so that the resultant Hilbert space depends neither on the initial spin ice configuration, nor the initial  spin flip. 
The ergodicity also takes care of the average over classical spin ice configurations in equation (\ref{eq:chiij2}).
Our 32-site cluster has periodic boundary conditions with lattice periods, $2\vec{a}_1, 2\vec{a}_2$, and $2\vec{a}_3$ [Fig.~\ref{Fig1} (a)]. To fully diagonalize the Hamiltonian (\ref{eq:MonopoleHamiltonian}), we consider  8 separate momentum sectors each of dimension 12348,  comfortably within the range of full diagonalization.

\begin{figure}[h]
\begin{center}
\includegraphics[width=0.45\textwidth]{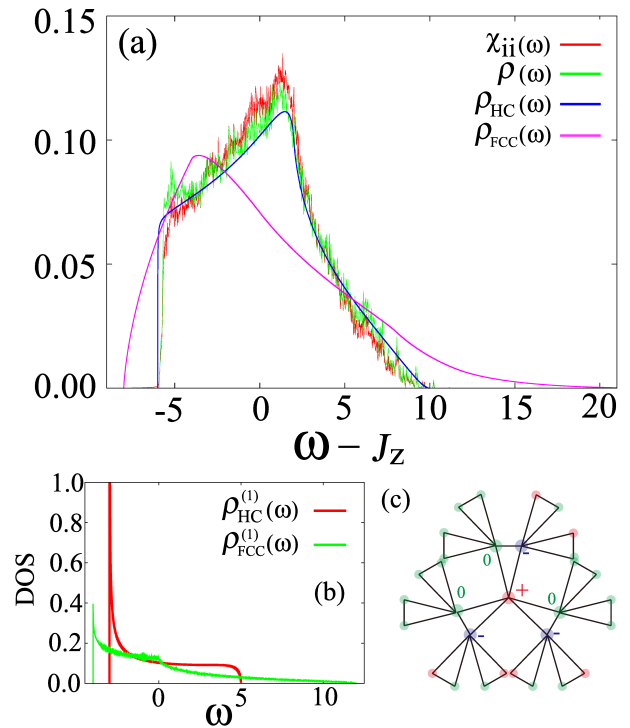}
\end{center}
\caption{\label{Fig2} 
(Color online) (a) $\chi_{ii}(\omega)$ and two-monopole density of states, $\rho(\omega)$, from  exact diagonalization of a 32-site cluster. $\rho(\omega)$ of the tight-binding model on Husimi cactus and FCC lattice are shown for comparison. (b) One-particle density of states on Husimi cactus and FCC lattice. (c) Schematic picture of the wave function at the lower spectral edge $\omega=-6$, depicted on the graph of many-body states.}
\end{figure}

The resulting local susceptibility, $\chi_{ii}(\omega)$ in Fig.~\ref{Fig2}(a) with $J_{\pm}/2=1$ as  energy unit, 
has a  highly asymmetric spectrum: 
a  steep rise at the low-energy spectral edge, $\omega=-6$, is followed 
by a peak around $\omega\sim1$ and a tail to higher energy. This asymmetry may be used to determine the sign
of $J_{\pm}$ in experiment, as flipping the sign of  $J_{\pm}$ amounts to inverting the x-axis, $\omega-J_z\to-(\omega-J_z)$.

The local susceptibility $\chi_{ii}(\omega)$ agrees remarkably well with the two-monopole DOS $\rho(\omega)$ [Fig.~\ref{Fig2}(a)].
Since  monopoles hop on the FCC lattice of tetrahedra, at first sight, one might expect 
the tight-binding spectrum of the FCC lattice to yield a useful approximation for $\rho(\omega)$.
However, we find that  the coupling to background spin ice changes the spectrum considerably.

To see this, consider the motion of a single monopole in detail.
As shown in Fig.~\ref{Fig1} (c), it can hop by flipping one of the three 
majority spins of the tetrahedron it hops from,
and the corresponding spin of a tetrahedron it hops to. 
By two further hops, the monopole can return to the initial tetrahedron.
Remarkably, after these three hops, not only the position of the monopole, but also the background spin configuration, 
remain unchanged.
A monopole can also return  to its initial location via a larger loop, Fig.~\ref{Fig1} (d).
However, in this case, the background spin configuration changes; it is only restored upon traversing the loop a second time.

These observations motivate us to formulate a hopping problem on 
the graph of many-body states. Each site of the graph represents a spin configuration
and a bond  connects  two sites  whenever $\mathcal{H}_{\rm eff}$ has a matrix element between the corresponding configurations. 
The monopole motion considered in Fig.~\ref{Fig1} (c) implies the existence of closed loops of length 3 on this graph. 
Omitting any further nontrivial closed loops -- as these are of the longer minimal length 6 --
the  graph in Fig.~\ref{Fig1} (b), known as Husimi cactus~\cite{doi:10.1063/1.1747725,0305-4470-27-5-019,PhysRevB.81.214445},  results.

This  turns out to work much better than the FCC analysis, as we show in Fig.~\ref{Fig2} (a):
the two-monopole DOS of the tight-binding model on the Husimi cactus quantitatively 
reproduces $\rho(\omega)$, and hence $\chi_{ii}(\omega)$.

The analysis of the Husimi cactus follows that of the motion of a mobile particle in an ice-rule potential 
on a simple Bethe lattice~\cite{PhysRevLett.104.226405}. The 
on-site Green's function 
$G(\varepsilon) = \frac{1}{\varepsilon - 6}\Bigl[\frac{3}{2}\sqrt{\frac{\varepsilon - 5}{\varepsilon + 3}} - \frac{1}{2}\Bigr]$
 gives the one-particle DOS
\begin{eqnarray}
\rho^{(1)}_{\rm HC}(\varepsilon)=\frac{3}{2\pi}\frac{1}{6-\varepsilon}\sqrt{\frac{5-\varepsilon}{3+\varepsilon}};
\label{eq:oneparticleDOS}
\end{eqnarray}
for details, see Supplemental material.
Fig.~\ref{Fig2}(b) shows $\rho^{(1)}_{\rm HC}$ alongside $\rho^{(1)}_{FCC}$ for the FCC lattice.

These two curves exhibit  crucial differences,  in (i) band width, and (ii) nature of the singularity at the lower band edge.
(i) The  Husimi cactus band width ($=8$) is halved compared with the FCC lattice ($=16$), due to the constraints imposed
on monopole hopping
by the background spin configuration, which allow flips only of majority spins. 
(ii) The lower-edge singularity, $\rho^{(1)}_{\rm FCC}(\varepsilon)$ is only 
a logarithmic divergence, $\propto-\log(\varepsilon-\varepsilon_{\rm min})$ with $\varepsilon_{\rm min}=-4$, the 
usual van-Hove singularity in three-dimensions. 
By contrast,  the onset at $\varepsilon_{\rm min}=-3$ 
for the Husimi cactus is more singular, $\propto(\varepsilon-\varepsilon_{\rm min})^{-1/2}$. 

Note that this square-root singularity in the density of states is that characteristic of a free particles in one dimension, even though the Husimi cactus, for which we have obtained this analytical result, is infinite dimensional in the same sense of the more familiar Bethe lattices/Cayley trees. At the same time, the physical motion of the monopoles actually takes place in three dimensional real space. This strikes us as notable in that the strong coupling of the monopoles to the gauge field background in spin ice leads to an effective dimensional transmutation, or perhaps more accurately, dimensional diversification.

The two-monopole DOS follow from the convolutions
\begin{eqnarray}
\rho_{\rm HC/FCC}(\omega) \equiv\int\ d\varepsilon\rho^{(1)}_{\rm HC/FCC}(\omega-\varepsilon)\rho^{(1)}_{\rm HC/FCC}(\varepsilon),
\label{eq:convolution}
\end{eqnarray}
plotted in [Fig.~\ref{Fig2} (a)].
{This  treats the monopoles are free particles, ignoring their interaction, as discussed below.}

$\rho_{\rm HC}(\omega)$ reproduces the two-monopole DOS of exact diagonalization to a remarkable accuracy.
This agreement implies several things.
Firstly, the result of exact diagonalization of the 32-site cluster
 is likely already a good approximation of thermodynamic limit, as the Husimi cactus calculation is not subject to finite-size effects.
Secondly, combined with the agreement of $\chi_{ii}(\omega)$ and two-monopole density of states, the analytic result also accurately
accounts for 
the experimentally observable q-integrated dynamical structure factor.

This suggests looking in experiment for the prominently singular edge structure of the spectrum, which corresponds to a step discontinuity.
It shows a steep rise at the band edge $\varepsilon=-6$, in  contrast to the two-particle DOS obtained from the FCC lattice via eq.~(\ref{eq:convolution}).
This reflects the stronger singularity of the one-particle DOS, $\rho^{(1)}_{\rm HC}(\varepsilon)\propto(\varepsilon-\varepsilon_{\rm min})^{-1/2}$.
For the size of the step, and hence the edge value of $\chi_{ii}(\omega)$, we obtain
\begin{eqnarray}
\rho_{\rm HC}(\omega\to-6^+)=\frac{2}{9\pi}\sim0.07077.
\end{eqnarray}

It  is even possible to obtain the one-monopole eigenfunction explicitly at this lower band edge.
The construction is analogous to the flat band of the tight-binding model on line graphs~\cite{0305-4470-25-16-011,PhysRevLett.69.1608}.
For its procedure, see for example, Ref~\cite{PhysRevB.81.014421}.  
On the graph shown in Fig.~\ref{Fig2} (c), the weight of eigenfunction $\psi_j$ at site $j$ is such that (a) $\psi_j=0$ or $\pm1$, and (b) on all the triangles, $\psi_j$ sums up to zero.
This construction gives an exact eigenstate of the tight-binding model on the Husimi cactus, with eigenenergy, $\varepsilon=-3$.
Mapping back to the original pyrochlore lattice, the corresponding many-body state describes the approximate one-monopole eigenstate of Hamiltonian (\ref{eq:MonopoleHamiltonian}),  given large loops are ignored.

\begin{figure}[h]
\begin{center}
\includegraphics[width=0.5\textwidth]{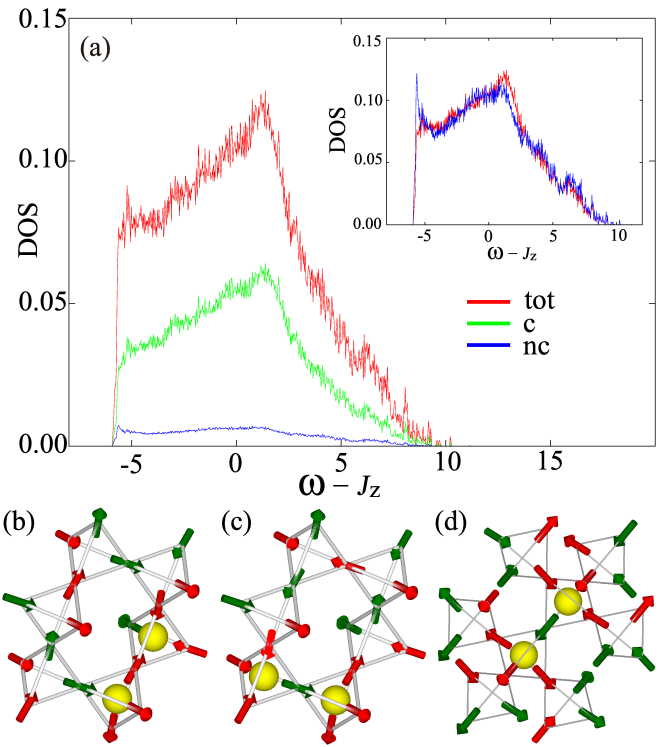}
\end{center}
\caption{\label{Fig3} 
(Color online) (a) Two-monopole density of states, $\rho(\omega)$, $\rho_{\rm c}(\omega)$, and $\rho_{\rm nc}(\omega)$. The inset shows the comparison between rescaled $\rho(\omega)$ and $\tilde{\rho}_{\rm nc}(\omega)$. (b) (c) Schematic picture of (b) contractible and (c) non-contractible monopole pairs. These two configurations are transformed to each other by encircling one monopole around the hexagonal ring. (d) A non-contractible pair on a loopless Husimi cactus. In this case, the pair cannot be deformed to a contractible pair.}
\end{figure}

We now turn to the effect of interactions between monopoles.
If monopoles are far apart, we can approximate their collective state as a direct product of the one-monopole states.
However, if they come closer -- and they do, as they are always pair created -- it is not possible to ignore their interactions.
In classical spin ice, these lead to non-trivial classical spin liquid phases~\cite{PhysRevLett.119.077207}, a liquid-gas phase transition~\cite{PhysRevLett.90.207205,castelnovo2008magnetic}, and collective phenomena in equilibrium and non-equilibrium settings~\cite{PhysRevB.94.104416,rau2016spin,PhysRevLett.119.077207}.
Here, we examine the pairing tendency of the monopoles.

Monopole encounters take two forms on a lattice, depending on whether the tetrahdra that host them share a minority or majority spin [Fig.~\ref{Fig3}(b) and (c)].
The former and the latter are called non-contractible and contractible pair, respectively~\cite{PhysRevLett.104.107201}.
Both situations can arise as components of the same eigenstate of $\mathcal{H}_{\rm eff}$, so that
\begin{eqnarray}
|m\rangle = a^{(m)}|\psi_m\rangle + a^{(m)}_{\rm nc}|\phi^{\rm nc}_m\rangle + a^{(m)}_{\rm c}|\phi^{\rm c}_m\rangle.
\end{eqnarray}
Here, $|\phi^{{\rm c}({\rm nc})}_m\rangle$ is the normalized vector composed only of the states with a (non)-contractible pair, 
and $|\psi_m\rangle$ contains the separated monopoles.
To examine the pairing tendency in different energy scales, we plot the pair-weighted two-monopole DOS,
\begin{eqnarray}
\rho_{{\rm c}({\rm nc})}(\omega)\equiv\sum_m|a^{(m)}_{{\rm c}({\rm nc})}|^2\delta(\omega - (J_z + \varepsilon_m)),
\end{eqnarray}
compared to the total two-monopole DOS in Fig.~\ref{Fig3} (a).

All three look similar overall.
For a detailed comparison, we use rescaled $\tilde{\rho}_{\rm nc}(\omega)=c\rho_{\rm nc}$, so that $\int\tilde{\rho}_{\rm nc}(\omega)d\omega=\int\rho(\omega)d\omega=1$, and compare $\tilde{\rho}_{\rm nc}(\omega)$ and $\rho(\omega)$ in the inset of Fig.~\ref{Fig3} (a).
There, we  find a spike for $\tilde{\rho}_{\rm nc}(\omega)$ at the low-energy edge.

In contrast, on the effectively loopless Husimi cactus,  the two types of monopole pairs are never connected, and
 two-monopole states thus define two separate sectors.
In the non-contractible sector, two monopoles are invisible to each other, Fig.~\ref{Fig3} (d). 
Accordingly, a two-monopole state in this sector can be expressed as a direct product of one-monopole states on the Husimi cactus,
with the result that the convolution formula (\ref{eq:convolution}) is exact, i.e. the two curves in the inset of Fig.~\ref{Fig3} (a) coincide 
perfectly.

The low-energy uprise of $\tilde{\rho}_{\rm nc}$, compared with $\rho(\omega)$, means that the loops of the pyrochlore lattice effect an  
attraction between monopoles for non-contractible pairs at low-energy. Such an attractive force is in principle interesting: 
in light of the possible softening of monopoles.
if the quantum exchange coupling, $J_{\pm}$, is sufficiently large, the system may eventually show an instability to a crystal phase 
involving non-contractible monopole pairs.

In summary, we have studied the quantum dynamics of gauge-charged fractional excitations in quantum spin ice.
We have identified the two-monopole DOS, $\rho(\omega)$, as a quantity which both is experimentally accessible and
exhibits features characteristic of the fractionalised setting. These include a marked asymmetry and a singular edge structure,
along with a spike related to interactions. 
We thus suggest extracting this quantity from inelastic neutron scattering data. 
These  features arise because of the rearrangement of the gauge
field degree of freedom, the `Dirac strings' attached to the monopoles~\cite{castelnovo2008magnetic}, which goes along with
monopole motion. From a methodological perspective, the success of our Husimi cactus treatment suggests that 
we have identified a setting in which the motion of an excitation on the graph of many-body states -- of autonomous interest in the 
separate context of e.g.\ many-body localisation~\cite{BASKO20061126,de2013support} -- appears to be a more natural description than that of motion in real space.

This work was supported by the JSPS KAKENHI (Nos. JP15H05852 and JP16H04026), MEXT, Japan, and by the Deutsche Forschungsgemeinschaft 
grant SFB1143. Part of
numerical calculations were carried out on the Supercomputer Center at Institute for Solid State Physics, University of Tokyo.
RM thanks Claudio Castelnovo, Olga Petrova and Shivaji Sondhi for collaboration on related work.

%

\newpage
\setcounter{figure}{0}
\setcounter{equation}{0}
\begin{center}
\Large 
{\it Supplemental information}
\end{center}
\section{Summary of Bethe lattice analysis}
\begin{figure}[h]
\begin{center}
\includegraphics[width=0.4\textwidth]{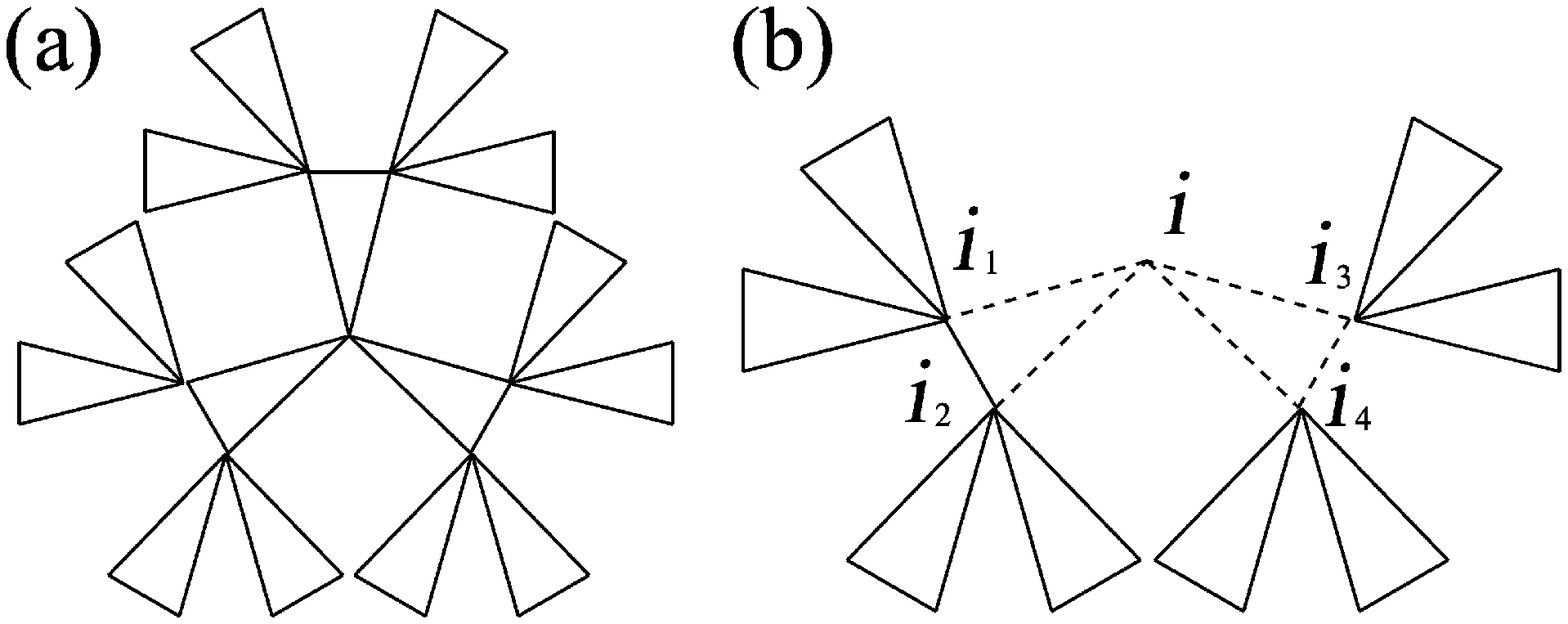}
\end{center}
\caption{\label{FigS1} 
(color online). (a) The three-leaf Bethe lattice. (b) A tree structure obtained from the three-leaf Bethe lattice by removing one of the branches extending from the site $i$.
If one further removes the two triangles shown with dashed lines, this network is decomposed into four self-similar networks extending from the sites, $i_1, i_2, i_3$ and $i_4$.
}
\end{figure}

Here, we introduce a derivation of the on-site Green's function on the three-leaf Bethe lattice [Fig.~\ref{FigS1} (a)], which leads to the one-particle DOS as shown in the equation (7) of the main text.
To this aim, we consider the tight-binding model on this network,
\begin{eqnarray}
\mathcal{H} = \sum_{\langle i,j\rangle}(|i\rangle\langle j| + {\rm H.c.}),
\end{eqnarray}
and define the on-site Green's function, 
\begin{eqnarray}
G_{i}(\varepsilon)\equiv\langle i|\frac{1}{\varepsilon - \mathcal{H} + i\delta}|i\rangle.
\end{eqnarray}

To obtain $G_{i}(\varepsilon)$, firstly, we cut off one of the branches extending from the site $i$, and obtain a self-similar network, as shown in Fig.~\ref{FigS1} (b).
Note that we obtain the four copies of the original network, if we further remove the two triangles involving the site $i$ (as shown with dashed lines in Fig.~\ref{FigS1} (b)).
We define the on-site Green's function, $g_i(\varepsilon)$, at site $i$ on this network.
By treating the hoppings on the two (dashed) triangles as a perturbation, one can construct a Dyson's equation,
\begin{align}
g_{i} = \frac{1}{\varepsilon} + 4\frac{1}{\varepsilon}g^2_{i}\frac{1}{1-g_{i}}.
\label{recursion1}
\end{align}
The Dyson's equation has a closed form, thanks to the self-similar nature of the network. It leads to a quadratic equation,
\begin{align}
(\varepsilon+4)g_i^2 - (\varepsilon+1)g_i + 1 =0.
\end{align}
By solving it, we obtain
\begin{align}
g_i = \frac{(\varepsilon+1) - \sqrt{(\varepsilon-5)(\varepsilon+3)}}{2(\varepsilon+4)} = \frac{2}{(\varepsilon+1) + \sqrt{(\varepsilon-5)(\varepsilon+3)}}.
\label{recursiveGreen}
\end{align}
Here, the sign before square root was chosen from the condition: $g_i(\varepsilon)\to\frac{1}{\varepsilon}$ as $\varepsilon\to\infty$.

\begin{figure}[h]
\begin{center}
\includegraphics[width=0.4\textwidth]{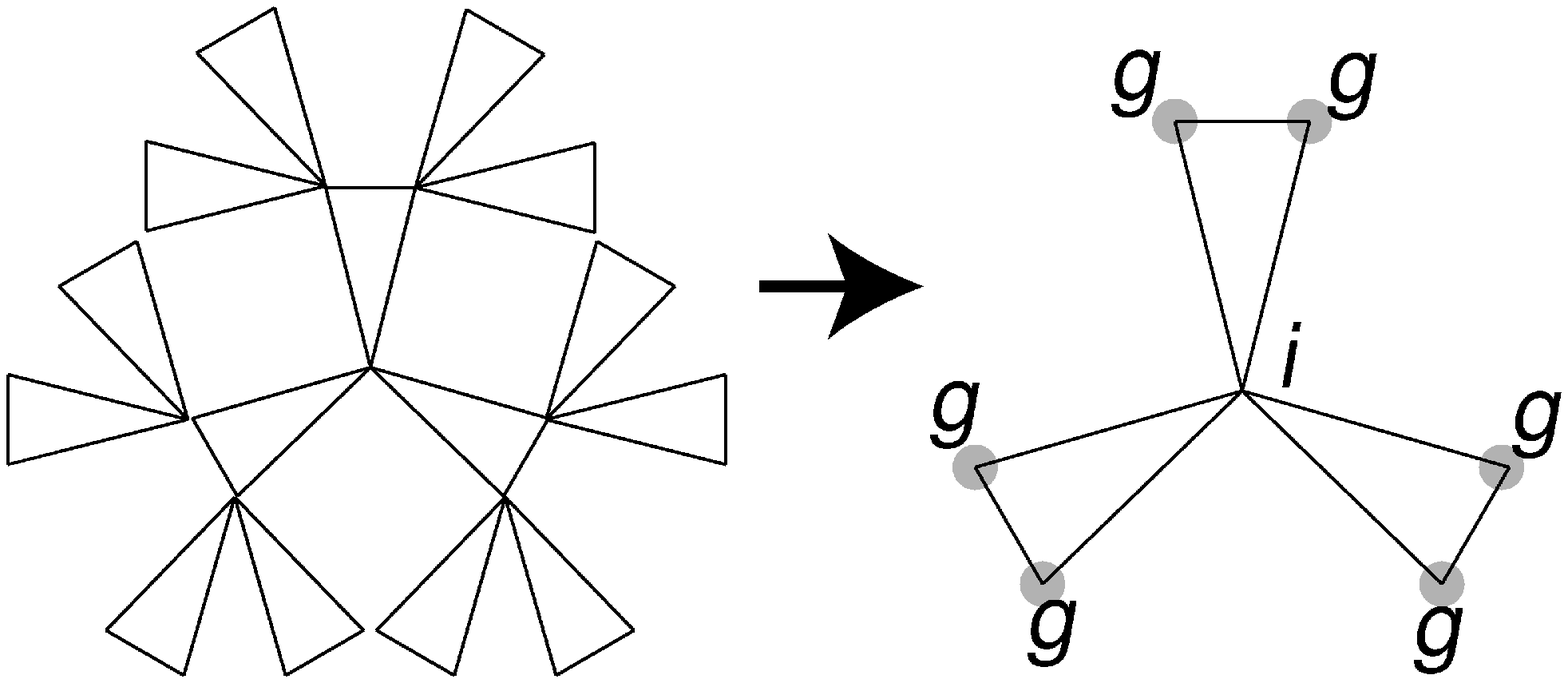}
\end{center}
\caption{\label{FigS2} 
(color online). The schematic picture to show that the hopping processes on the branches can be taken into account as the on-site Green's function.
}
\end{figure}

Next, we obtain the on-site Green's function on the original three-leaf Bethe lattice, $G_i(\varepsilon)$.
To this aim, we again treat the hoppings on the triangles involving the site $i$, as a perturbation. 
To construct a Dyson's equation, it is convenient to replace the whole network with the ``clover" as shown in Fig.~\ref{FigS2}, 
where the on-site Green's functions at the outer sites are renormalized to be $g_i(\varepsilon)$, taking account of the contributions from the branches extending from these sites.
Then, the Dyson's equation reads
\begin{align}
G_i(\varepsilon) = \frac{1}{\varepsilon} + 6\frac{1}{\varepsilon}\frac{g_i}{1-g_i}G_i.
\end{align}
By solving it, we obtain
\begin{align}
G_i(\varepsilon) = \frac{1}{\varepsilon - 6}\Bigl[\frac{3}{2}\sqrt{\frac{\varepsilon - 5}{\varepsilon + 3}} - \frac{1}{2}\Bigr].
\label{onsiteG}
\end{align}
This equation leads to the one-particle density of states shown in equation (7) of the main text.
Note that $G_i(\varepsilon)$ has an isolated pole at $\varepsilon=6$, however, the delta-functional peak is missing at this energy due to the vanishing residue at this pole,
with the result that only the continuum spectrum remains for $-3<\varepsilon<5$.

\end{document}